# On Determining the Fair Bandwidth Share for ABR Connections in ATM Networks[*]

Sonia Fahmy
Department of Computer Sciences, Purdue University, E-mail: fahmy@cs.purdue.edu

Raj Jain
Department of CIS, The Ohio State University, E-mail: jain@cis.ohio-state.edu

Shivkumar Kalyanaraman
Department of ECSE, Rensselaer Polytechnic Institute, E-mail: shivkuma@ecse.rpi.edu

Rohit Goyal
Axiowave Networks, E-mail: rgoyal@axiowave.com

Bobby Vandalore
Nokia, E-mail: bobby.vandalore@nokia.com

**Abstract:** In a multi-service network such as ATM, adaptive data services (such as ABR) share the bandwidth left unused by higher priority services. The network indicates to the ABR sources the *fair and efficient* rates at which they should transmit to minimize their cell loss. Switches must constantly measure the demand and available capacity, and divide the capacity fairly among the contending connections. In this paper, we propose a new method for determining the "effective" number of active connections, and the fair bandwidth share for each connection. We prove the efficiency and fairness of the proposed method analytically, and use several simulations to illustrate its fairness dynamics and transient response properties.

# 1 Introduction

ATM networks offer five service categories: constant bit rate (CBR), real-time variable bit rate (rt-VBR), non-real time variable bit rate (nrt-VBR), available bit rate (ABR), and unspecified bit rate (UBR). The ABR and UBR service categories are specifically designed for data traffic. The ABR service provides better service for data traffic than UBR by frequently indicating to the sources the rate at which they should be transmitting to minimize loss. For this reason, an ATM switch must continuously compute a fair and efficient bandwidth share for each of the currently active ABR connections.

---

[*]This paper is an extended version of paper [7] presented at the IEEE International Conference on Communications (ICC) 1998.

Determining the fair bandwidth share for the active ABR connections is a complex problem. A number of fairness objectives, including max-min fairness and proportional fairness have been proposed. Intuitively, max-min fairness means that if a connection is bottlenecked elsewhere, it should be allocated the maximum it can use at this switch, and the left over capacity should be fairly divided among the connections that can use it. The switch should indicate this fair bandwidth share to the sources, while also accounting for the load and queuing delays at the switch.

This paper proposes a novel method to determine the fair bandwidth share for the active ABR connections, and analyzes the performance of this method using both simple mathematical proofs and simulations. The remainder of the paper is organized as follows. In the next section, we review the ABR flow control mechanisms in ATM networks. Then, we describe the original ERICA switch algorithm [14] which is employed in this study as a framework on which to develop the new method. Sections 4 and 5 point out some problems with the original ERICA algorithm, and describe how ERICA has solved these problems. We then describe our proposed method (which also overcomes those problems), and give a proof of its correctness, and a number of examples of its operation. Finally, we analyze the performance of the proposed method and compare it to ERICA.

## 2   The ABR Flow Control Mechanism

As previously mentioned, the ABR service frequently indicates to the sources the rate at which they should be transmitting. The switches monitor their load, compute the available bandwidth and divide it fairly among the active flows. The feedback from the switches to the sources is indicated in Resource Management (RM) cells which are generated periodically by the sources and turned around by the destinations. Figure 1 illustrates this operation.

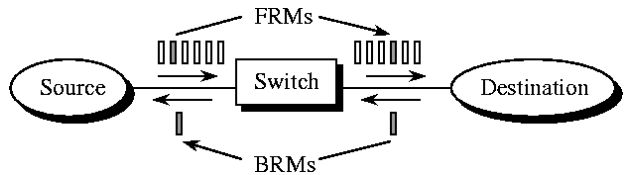

Figure 1: Resource management cells in an ATM network

The RM cells contain the source current cell rate (CCR), in addition to several fields that can be used by the switches to provide feedback to the sources. These fields are: the explicit rate (ER), the congestion indication (CI) flag, and the no increase (NI) flag. The ER field indicates the rate that the network can support for this connection at that particular instant. At the source, the ER field is initialized to a rate no greater than the PCR (peak cell rate), and the CI and NI flags are usually reset. On the path, each switch reduces the ER field to the maximum rate it can support, and sets CI or NI if necessary [9].

The RM cells flowing from the source to the destination are called forward RM cells (FRMs) while those returning from the destination to the source are called backward RM cells (BRMs) (see figure 1). When a source receives a BRM cell, it computes its allowed cell rate (ACR) using its current ACR value, the CI and NI flags, and the ER field of the RM cell [11].

## 2.1 Fairness Criterion

The optimal operation of a distributed shared resource, bandwidth in our case, is given by a criterion called the *max-min allocation* [8, 2, 15]. This fairness definition is the most intuitive, though proportional sharing has recently gained significant attention. Max-min allocation gives equal shares to sources bottlenecked at the same link, utilizing all capacity left over by non-bottlenecked sources. Given a configuration with $n$ contending sources, suppose the $i^{th}$ source is allocated a bandwidth $x_i$. The allocation vector $\{x_1, x_2, \ldots, x_n\}$ is feasible if all link load levels are less than or equal to 100%. Given an allocation vector, the source with the least allocation is, in some sense, the "unhappiest source." We find the feasible vectors that give the maximum allocation to this unhappiest source (thus maximizing the minimum source, or max-min). Then, we remove this "unhappiest source" and reduce the problem to that of the remaining $n-1$ sources operating on a network with reduced link capacities. We repeat this process until all sources have been allocated the maximum that they can obtain.

# 3  The Original ERICA Switch Algorithm

Several switch algorithms have been developed to compute the feedback to be indicated to ABR sources in RM cells [1, 13, 16, 17, 12]. The ERICA algorithm [10, 12] is one of the earliest and most studied explicit rate algorithms. The main advantages of ERICA are its low complexity, fast transient response, efficient allocations, and controlled queuing delay.

In this section, we present the basic features of the original algorithm and explain their operation. The next sections describe an addition to the basic algorithm, and a new alternative method to determine the number of active connections. For a more complete description of the algorithm and its performance, refer to [12].

The ERICA switch periodically monitors the load on each link and determines a load factor, $\rho$, the available capacity, and the number of currently active virtual connections (VCs). The load factor is calculated as follows:

$$\rho \leftarrow \frac{\text{ABR Input Rate}}{\text{ABR Capacity}}$$

where:
ABR Capacity $\leftarrow$ Target Utilization $\times$ Link Bandwidth $-$ VBR Usage $-$ CBR Usage.

The input rate and output link ABR capacity are measured over an interval called the switch measurement interval. The above steps are executed at the end of the switch measurement interval. Target utilization is a parameter which is set to a fraction based on current queuing delay. The load factor, $\rho$, is an indicator of the congestion level of the link. The optimal operating point is at $\rho$ close to one.

The fair share of each VC, $FairShare$, is also computed as follows:

$$\text{FairShare} \leftarrow \frac{\text{ABR Capacity}}{\text{Number of Active Connections}}$$

The switch allows each connection sending at a rate below the $FairShare$ to rise to $FairShare$. If the connection does not use all of its $FairShare$, then the switch fairly allocates the remaining capacity to the connections which can use it. For this purpose, the switch calculates the quantity:

$$\text{VCShare} \leftarrow \frac{CCR}{\rho}$$

If all VCs changed their rate to their $VCShare$ values, then, in the next cycle, the switch would experience unit load ($\rho = 1$). $VCShare$ aims at bringing the system to an efficient operating point, which may not necessarily be fair. A combination of the $VCShare$ and $FairShare$ quantities is used to rapidly reach optimal operation as follows:

$$\text{ER Calculated} \leftarrow \text{Max (FairShare, VCShare)}$$

The calculated ER value cannot be greater than the ABR Capacity which has been measured earlier. Hence, we have:

$$\text{ER Calculated} \leftarrow \text{Min (ER Calculated, ABR Capacity)}$$

To ensure that the bottleneck ER reaches the source, each switch computes the minimum of the ER it has calculated as above and the ER value in the RM cell, and indicates this value in the ER field of the RM cell.

The algorithm described above is the basic algorithm, but several other steps are carried out to avoid transient overloads and variations in measurement, and drain the transient queues. Moreover, the algorithm is modified to achieve max-min fairness as described in sections 5 and 6.

## 4  The Measurement Interval

ERICA measures the required quantities over consecutive intervals and uses measured quantities in each interval to calculate the feedback in the next interval. The length of the measurement interval limits the amount of variation which can be eliminated. It also determines how quickly the feedback can be given to the sources, because ERICA gives the same feedback value per source during each measurement interval. Longer intervals produce better averages, but slow down the rate of feedback. Shorter intervals may result in more variation in measurements, and may consistently underestimate or overestimate the measured quantities.

The ERICA algorithm estimates the number of active VCs to use in the computation of the fair share by considering a connection active if the source sends *at least one cell during the measurement interval.* This can be **inaccurate** if the source is sending at a low rate and the measurement interval is short. Exponentially averaging the number of active connections over successive intervals produces more accurate estimates, but may still underestimate the number of connections if the measurement interval is short. In this paper, we propose a more accurate method for estimating connection activity. The new method is insensitive to the length of the measurement interval. It also eliminates the need to perform some of the steps of the ERICA algorithm, as described in the next section.

## 5  The Fairness Problem and ERICA Solution

Assuming that measurements do not exhibit extremely high variation, the original ERICA algorithm converges to efficient operation in all cases. The convergence from transient conditions to the desired operating point is rapid, often taking less than a round trip time. We have, however, discovered cases in which the original algorithm does not converge to max-min fair allocations. This happens if all of the following three conditions are met: (1) the load factor $\rho$ becomes one, (2) there are some connections which are bottlenecked upstream of the switch under consideration, (3) the source rate for all remaining connections is greater than the $FairShare$. In this case, the system remains in its current state, because the term $CCR/\rho$ is greater than $FairShare$ for the non-bottlenecked connections.

This problem was overcome in ERICA as follows. The algorithm is extended to remember the highest allocation made during each measurement interval, and ensure that all eligible connections can also get this same high allocation. To do this, $MaxAllocPrevious$ stores the maximum allocation given in the previous interval. For $\rho > 1 + \delta$, where $\delta$ is a small fraction, we use the basic ERICA algorithm and allocate Max (FairShare, VCShare). But, for $\rho \leq 1+\delta$, we attempt to make all the rate allocations equal, by assigning ER to Max (FairShare, VCShare, MaxAllocPrevious). The aim of introducing the quantity $\delta$ is to force the allocation of equal rates when the overload is fluctuating around unity, thus avoiding unnecessary rate oscillations. The remainder of this paper proposes a more accurate method to compute the max-min fair shares for all the contending connections.

## 6  An Accurate Method to Determine the Fair Bandwidth Share

As previously discussed, ERICA determines the number of active connections by considering a source as active if at least one cell from this source is sent during the measurement interval. A more accurate method to compute activity and eliminate the need for the proposed solution to the fairness problem is to compute a quantity that we call the "effective number of active VCs" and use this quantity to compute the $FairShare$, as described next.

### 6.1  Basic Idea

We redefine the $FairShare$ quantity to be ***the maximum share a VC could get at this switch under max-min fairness criteria***. Hence, the $FairShare$ is calculated as follows:

$$\text{FairShare} = \frac{\text{ABR capacity}}{\text{Effective number of active VCs}}$$

The main innovation is the computation of the effective number of active VCs. The value of the effective number of active VCs depends on the activity level of each of the VCs. The activity level of a VC is defined as follows:

$$\text{Activity level} = \text{Min}(1, \frac{\text{Source Rate}}{FairShare})$$

Thus, VCs that are operating at or above the $FairShare$ are each counted as one. The VCs that are operating below the $FairShare$ (because they are not bottlenecked at this switch, or because they are variable demand applications) only contribute a fraction. The VCs that are bottlenecked at this switch are considered fully active while other VCs are considered partially active.

The effective number of active VCs is the sum of the activity levels for all VCs:

$$\text{Effective number of active VCs} = \sum_i \text{Activity level of VC}_i$$

Note that the definition of activity level depends upon the $FairShare$, and the definition of the $FairShare$ depends upon the activity levels. Thus, the definitions are recursive. Ideally, we would need to iterate several times given the rates of various VCs.

## 6.2 Examples of Operation

**Example 1 (stability):**

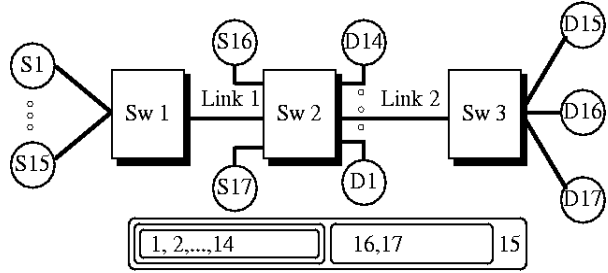

Figure 2: Upstream Configuration

Consider the upstream bottleneck case with 17 VCs shown in figure 2. We have shown that this configuration demonstrates the unfairness of the original ERICA algorithm as described in section 3, which necessitates the addition described in section 5.

Assume that the target capacity is 150 Mbps. For the second switch, when the rates for ($S1$, $S16$, $S17$) are (10, 70, 70):

>Iteration 1: FairShare = 70 Mbps
>>Activity = (10/70, 70/70, 70/70) = (1/7, 1, 1)
>>Effective number of active VCs = 1 + 1 + 1/7 = 15/7
>
>Iteration 2: FairShare = Target capacity/Effective number of active VCs = 150/2.14 = approximately 70 Mbps

Hence, this example shows that the system is stable at the allocation of (10, 70, 70). At any other allocation, the scheme will calculate the appropriate $FairShare$ that makes the allocation eventually reach this point, as seen in the next two examples.

**Example 2 (rising from a low FairShare):**

For the same configuration, when the rates are (10, 50, 90):

>Assume that the Effective number of active VCs = 3
>Iteration 1: FairShare = 150/3 = 50 Mbps
>>Activity = (10/50, 50/50, 1) = (0.2, 1, 1)
>>Effective number of active VCs = 0.2 + 1 + 1 = 2.2
>
>Iteration 2: FairShare = 150/2.2 = approximately 70 Mbps

Again, the scheme reaches the optimal allocation within a few round trip times.

**Example 3 (dropping from a high FairShare):**

For the same configuration, when the rates are (10, 50, 90), suppose that the effective number of active VCs is initially 2:

>Iteration 1: FairShare = 150/2 = 75 Mbps
>>Activity = (10/75, 50/75, 1) = (0.13, 0.67, 1)
>>Effective number of active VCs = 0.13 + 0.67 + 1 = 1.8

Iteration 2: FairShare = 150/1.8 = 83.33 Mbps

Suppose the sources start sending at the new rates, except for the first one which is bottlenecked at 10 Mbps. Also assume that FairShare is still at 83.33 Mbps.

Activity = (10/83.33, 83.33/83.33, 83.33/83.33) = (0.12, 1, 1)
Effective number of active VCs = 0.12 + 1 + 1 = 2.12
FairShare = 150/2.12 = approximately 70 Mbps

Again, the scheme reaches the optimal allocation after the sources start sending at the specified allocations, which is within a few round trip times.

## 6.3 Derivation

The following derivation shows how we have verified the correctness of our method of calculation of the number of active connections. The new algorithm is based upon some of the ideas presented in the MIT scheme [3, 4, 5]. However, this algorithm does not suffer from the known drawbacks of the MIT scheme: its high complexity, possible underutilization, and insensitivity to queuing delay.

The derivation depends on classifying active VCs as either underloading VCs or overloading VCs. A VC is *overloading* if it is bottlenecked at this switch; otherwise the VC is said to be *underloading*. In the MIT scheme, a VC is determined to be overloading by comparing the computed $FairShare$ value to the desired rate indicated by the VC source. In our scheme, we classify a VC as overloading if its source rate is greater than the $FairShare$ value. Our algorithm only performs one iteration every measurement interval ($O(1)$), and is not of the complexity of the order of the number of VCs ($O(N)$), as with the MIT scheme.

The MIT scheme has been proven to compute max-min fair allocations for connections within a certain number of round trips (see the proof in [4]). According to the MIT scheme:

$$FairShare = \frac{\text{ABR Capacity} - \sum_{i=1}^{\mathcal{N}_u} \mathcal{R}u_i}{\mathcal{N} - \mathcal{N}_u}$$

where:
$\mathcal{R}u_i$ = Rate of $i^{th}$ underloading source ($1 \leq i \leq \mathcal{N}_u$)
$\mathcal{N}$ = Total number of VCs
$\mathcal{N}_u$ = Number of underloading VCs

Substituting $\mathcal{N}_o$ for the denominator term, this becomes:

$$\text{FairShare} = \frac{\text{ABR Capacity} - \sum_{i=1}^{\mathcal{N}_u} \mathcal{R}u_i}{\mathcal{N}_o}$$

where:
$\mathcal{N}_o$ = Number of overloading VCs ($\mathcal{N}_u + \mathcal{N}_o = \mathcal{N}$)

Multiplying both sides by $\mathcal{N}_o$, we get:

$$FairShare \times \mathcal{N}_o = \text{ABR Capacity} - \sum_{i=1}^{\mathcal{N}_u} \mathcal{R}u_i$$

Adding $\sum_{i=1}^{\mathcal{N}_u} \mathcal{R}u_i$ to both sides produces:

$$FairShare \times \mathcal{N}_o + \sum_{i=1}^{\mathcal{N}_u} \mathcal{R}u_i = \text{ABR Capacity}$$

Factoring $FairShare$ out in the left hand side:

$$FairShare \times (\mathcal{N}_o + \sum_{i=1}^{\mathcal{N}_u} \frac{\mathcal{R}u_i}{FairShare}) = \text{ABR Capacity}$$

Or:
$$FairShare = \frac{\text{ABR Capacity}}{\mathcal{N}_o + \sum_{i=1}^{\mathcal{N}_u} \frac{\mathcal{R}u_i}{FairShare}}$$

Substituting $\mathcal{N}_{eff}$, we get:
$$FairShare = \frac{\text{ABR Capacity}}{\mathcal{N}_{eff}}$$

where:
$$\mathcal{N}_{eff} = \mathcal{N}_o + \sum_{i=1}^{\mathcal{N}_u} \frac{\mathcal{R}u_i}{FairShare}$$

This means that the effective number of active VCs is equal to the number of overloading sources, plus the fractional activity of underloading sources. This is the key equation we have proposed above, and implemented as discussed in the next subsection.

### 6.4 Algorithm Pseudo-code

This section explains how the new algorithm was implemented and incorporated into the ERICA switch algorithm. The following variables are introduced:

- $N_{last}$: Effective number of active VCs in the last measurement interval.

- $N_{current}$: Effective number of active VCs being accumulated for the current measurement interval.

- Activity: This array is maintained for each VC. It is set to one for overloading sources (an overloading source is a source whose CCR exceeds its $FairShare$ value). The activity of a VC is set to the fraction obtained from dividing the CCR of the VC by the $FairShare$ value in the case of underloading sources.

- FirstCellSeen: This is also maintained for each VC, and is only used to avoid the initialization effects of the VC. It is one bit that is set to one if the VC has shown any sign of activity; otherwise, it is set to zero.

- VCsSeen: The sum of the VCs whose FirstCellSeen flag is set. Also used to avoid initialization effects.

**INITIALIZATION:**

1. $N_{last}$ = number of VCs set up

2. $FairShare$ = ABR Capacity/$N_{last}$

3. $N_{current}$ = 0

4. VCsSeen = 0

5. FOR ALL VCs DO
   Activity [VC] = 0
   FirstCellSeen [VC] = 0
   END (* FOR *)

6. Initialize other ERICA variables

**END OF MEASUREMENT INTERVAL:**

1. IF (VCsSeen >= $N_{last}$)
   $N_{last}$ = max (1, $N_{current}$)
   END (* IF *)

2. $N_{current}$ = 0

3. $FairShare$ = ABR Capacity/$N_{last}$

4. FOR ALL VCs DO
   Activity [VC] = min (1, CCR [VC]/$FairShare$)
   $N_{current}$ = $N_{current}$ + Activity [VC]
   END (* FOR *)

5. Update Overload Factor, and update or reset other ERICA variables

**CELL IS RECEIVED IN FORWARD DIRECTION:**

1. Do NOT update $N_{current}$ as used to be done with ERICA

2. IF (NOT FirstCellSeen [VC]) THEN
   FirstCellSeen [VC] = 1
   VCsSeen = VCsSeen + 1
   END (* IF *)

3. Update CCR [VC]

**BRM CELL TO BE SENT IN REVERSE DIRECTION:**
ER Computed = Max (FairShare, CCR [VC]/Overload Factor)

Observe that the FirstCellSeen array and the VCsSeen counter are only used for the purpose of removing initialization effects from the simulation, and will not exist in a real implementation. Thus, in a real implementation, no steps (other than source rate estimation) will be carried out when a cell is seen, which means that the algorithm will have a low complexity.

# 7 Performance Analysis

The new algorithm has been tested for a variety of networking configurations using several performance metrics. The results were similar to the results obtained with the ERICA algorithm [10], except that the new algorithm is max-min fair (without executing the max-min fairness step described in section 5 above), and also the algorithm is less sensitive to the length of the measurement interval. A sample of the results is discussed in this section.

## 7.1 Parameter Settings

Throughout our experiments, the following parameter values are used:

1. All links have a bandwidth of 155.52 Mbps.

2. All links are 1000 km long.

3. All VCs are bidirectional.

4. The source parameter Rate Increase Factor (RIF) is set to one, to allow immediate use of the full explicit rate indicated in the returning RM cells at the source.

5. The source parameter Transient Buffer Exposure (TBE) is set to large values to prevent rate decreases due to the triggering of the source open-loop congestion control mechanism. This was done to isolate the rate reductions due to the switch congestion control from the rate reductions due to TBE.

6. The switch target utilization parameter was set to 90%. This factor is used to scale down the ABR capacity term used in the ERICA algorithm. Alternatively, it can be dynamically computed based upon the current queuing delay at the switch.

7. The switch measurement interval was set to the minimum of the time to receive 100 cells and 1 ms.

8. All sources are deterministic, i.e., their start/stop times and their transmission rates are known. Hence, we did not need to conduct several simulation runs (with different random number generator seeds) and average the results.

## 7.2 Simulation Results

The simulations performed focus on two main aspects of the new scheme: its fairness, and its transient response.

### 7.2.1 Fairness

In order to test the fairness of the new algorithm, we simulated a three source configuration where one of the sources is bottlenecked at a low rate (10 Mbps). Hence, even though the network gives that source feedback to increase its rate, it never sends at a rate faster than 10 Mbps. The other two sources start transmission at different initial rate (ICR) values. The aim of this configuration is to examine whether the two non-bottlenecked sources will reach the same ACR values, utilizing the bandwidth left over by the first source.

Figure 3 illustrates the topology of the configuration simulated. Note that the round trip time for the $S2$ and $S3$ connections is 30 ms, while that for the $S1$ connection is 40 ms. This configuration is almost identical to the one used in the examples in section 6 (figure 2), except that connection $S1$ to $D1$ is bottlenecked at the source $S1$ itself, and not at "Link 1." The reason we chose to demonstrate a source bottleneck situation here (and not a link bottleneck situation like figure 2) is to demonstrate the effect of using the CCR field in the RM cells versus measuring the source rate.

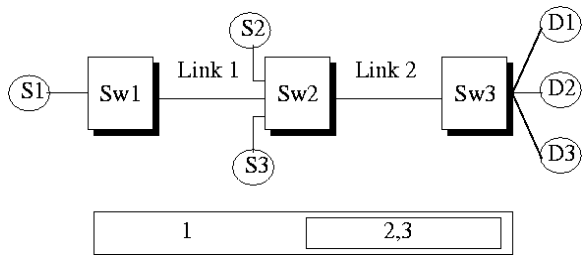

Figure 3: Three source configuration

The results are presented in the form of three graphs for each configuration:

1. Graph of allowed cell rate (ACR) in Mbps over time for each source.

2. Graph of ABR queue lengths in cells over time at the bottleneck port.

3. Graph of the effective number of active VCs $N_{eff}$ at the bottleneck port.

Figure 4 illustrates the performance of the original ERICA algorithm without the fairness step discussed in section 5. Source $S1$ is the bottlenecked source. Sources $S2$ and $S3$ start sending at different ICR (and hence ACR) values. Their ICR values and that of $S1$ add up to little more than the the link rate, so the initial $\rho$ value at the switch is almost one. Observe that the rates of $S2$ and $S3$ remain different, leading to unfairness. The number of active VCs is determined using the original ERICA method, so the switch sees 3 sources (see figure 4(c)), and the $FairShare$ value remains at around 50 Mbps. Hence, the source $S2$ never increases its rate to make use of the bandwidth left over by $S1$ and only $S3$ utilizes this bandwidth.

Figure 5 illustrates how the fairness problem was overcome in ERICA by the change described in section 5. In this case, the sources are given the maximum allocation in case of underload or unit load, and hence all sources get an equal allocation. The modified algorithm is max-min fair.

Figure 6 illustrates the results with the new method to calculate the fair share of the bandwidth. Observe that the allocations are max-min fair in this case, without needing to apply the maximum allocation algorithm as in the previous case. This is because the new method used to compute the "effective" number of active connections is used. Figure 6 shows that after the initialization period, the effective number of active VCs stabilizes at 1 (for $S2$), plus 1 (for $S3$), plus 10/50 (for $S1$), which gives $1 + 1 + 0.2 = 2.2$ sources. The method also stabilizes to the correct value even *if the length of the measurement interval is short*, unlike the original method where the length of the measurement interval must be long enough to detect cells from all sources, even low-rate sources.

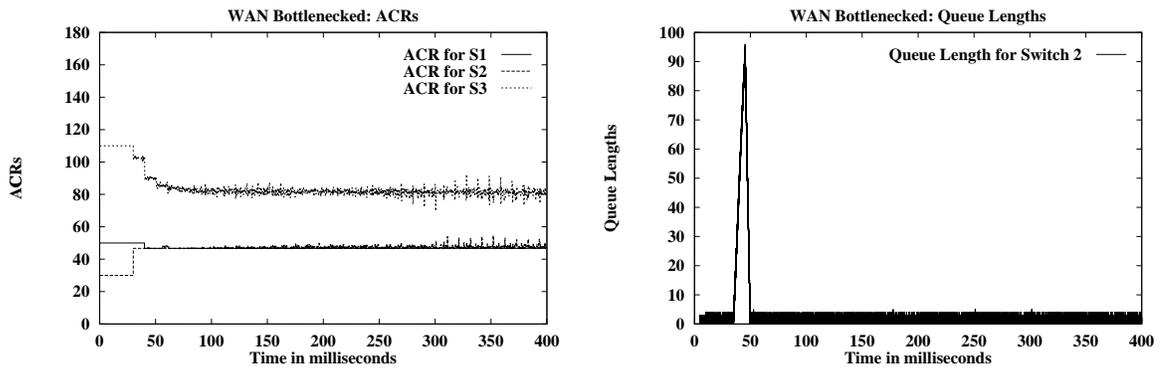

(a) Allowed Cell Rate

(b) Queue Length

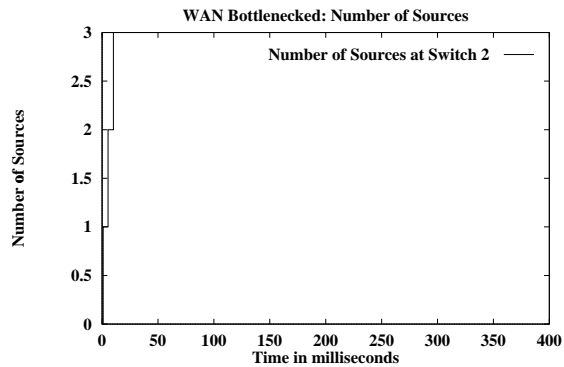

(c) Number of Active Connections

Figure 4: Results for a WAN three source bottleneck configuration with the original ERICA

The proposed method works correctly for all cases when there are *link bottlenecks* at various locations (e.g., the configuration in figure 2), since it correctly computes the activity level of each connection based on its CCR value. However, observe that in *source bottleneck* cases, the CCR value cannot be simply obtained from the forward RM cells, but must be measured by the switches. This is because, in source bottleneck situations, the source indicates its ACR value in the CCR field of the RM cell, but the source may actually be sending at a much lower rate than its ACR.

For example, for the configuration discussed above (figure 3), assume that we were relying on the CCR values in the RM cells. Figure 7 shows that the new method is not fair in this case, since source $S1$ indicates an ACR of 50 Mbps so the effective number of active connections stabilizes at 3 (see figure 7(c)), and the $FairShare$ remains at 50 Mbps. But source $S1$ is only sending at 10 Mbps. CCR measurement at the switch detects this, and hence arrives at the correct allocation as seen in figure 6.

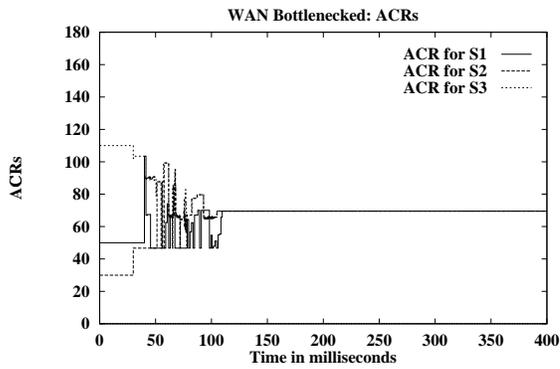

(a) Allowed Cell Rate

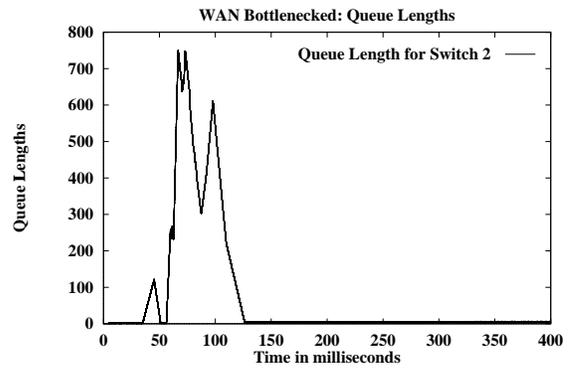

(b) Queue Length

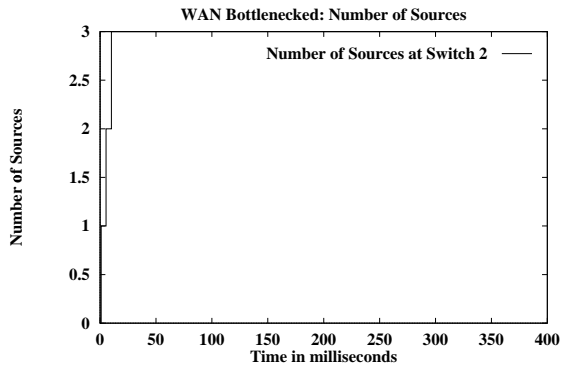

(c) Number of Active Connections

Figure 5: Results for a WAN three source bottleneck configuration with ERICA

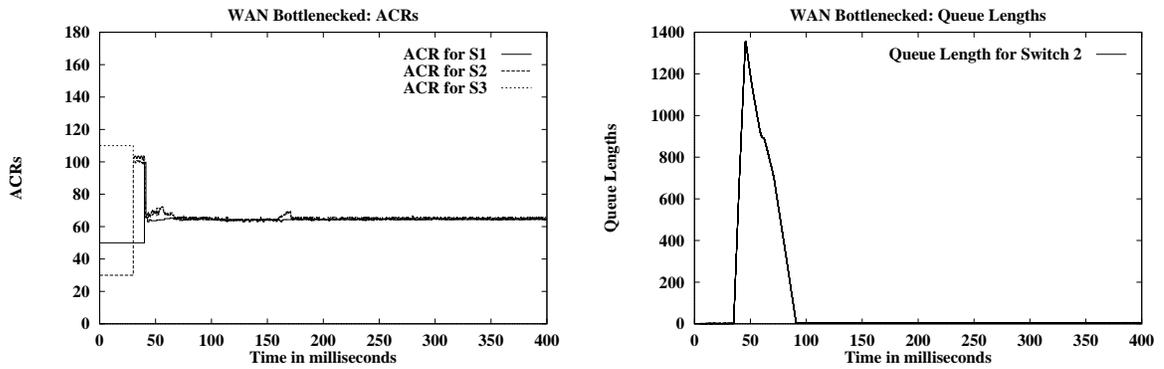

(a) Allowed Cell Rate

(b) Queue Length

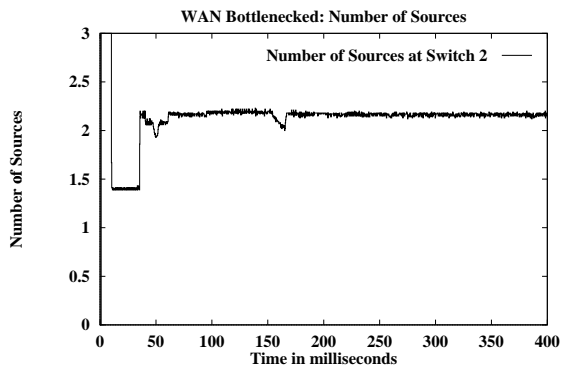

(c) Number of Active Connections

Figure 6: Results for a WAN three source bottleneck configuration with the proposed ERICA and source rate measurement at the switch

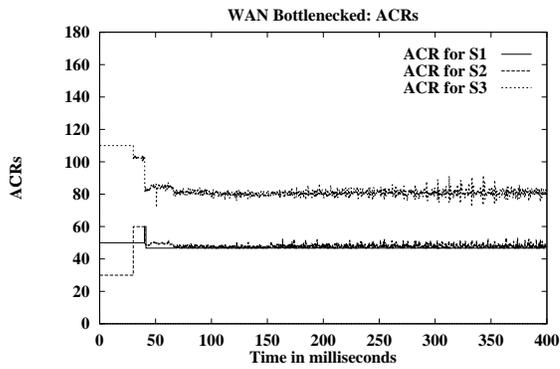

(a) Allowed Cell Rate

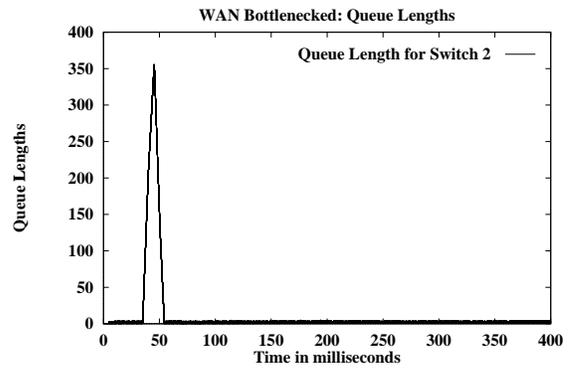

(b) Queue Length

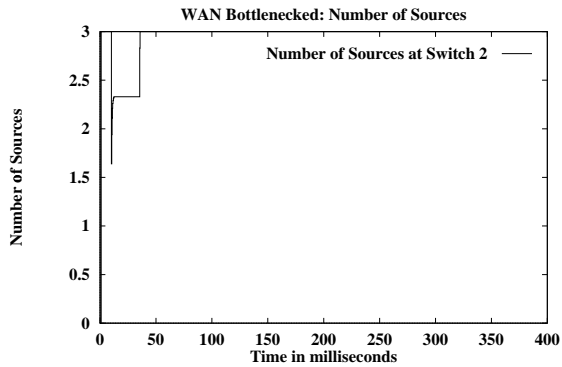

(c) Number of Active Connections

Figure 7: Results for a WAN three source bottleneck configuration with the proposed ERICA

### 7.2.2 Transient Response

One of the main properties of the new algorithm is that, unlike the MIT scheme, it is $O(1)$. Due to this, more than one round trip time may be required to arrive at the optimal allocations. In order to determine if the proposed method has a significantly slower transient response due to its recursive operation, we run another set of simulations.

Figure 8 illustrates the two-source configuration we used in this set of simulations. The round trip time for each connection is 30 ms. The new algorithm was simulated for this configuration, where the first source is active throughout the simulation period, while the second source starts sending after 60 ms and stops sending data at 120 ms. Both sources are persistent sources while they are active.

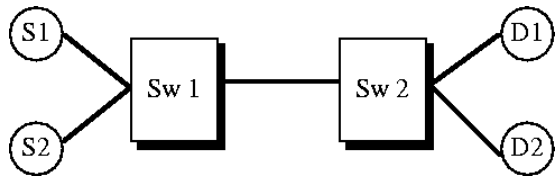

Figure 8: Two source configuration

The results are presented in the form of two graphs for each configuration:

1. Graph of allowed cell rate (ACR) in Mbps over time for each source.

2. Graph of link utilization (as a percentage) over time for the bottleneck link.

Figures 9 and 10 show the performance of the ERICA algorithm (with the fairness modification) versus the performance of the proposed algorithm. It is clear that the transient response of both methods is comparable. The new method is slightly slower in reducing the rates in the start-up period of the second source, due to the recursive nature of the algorithm. However, the difference is small, and the benefits of the method outweigh the slower response.

### 7.3 Observations on the Results

From the simulation results, we can make the following observations about the performance of the proposed algorithm:

- During transient phases, if the $FairShare$ value increases, the $N_{eff}$ value decreases (since it uses the $FairShare$ value in the denominator), and $FairShare$ further increases (since it uses $N_{eff}$ in the denominator), so $N_{eff}$ further decreases, and so on, until the correct values of source rate, $N_{eff}$ and $FairShare$ are reached. Then the proposed scheme is provably fair and efficient in steady state (see figure 6(a) and (c)).

- Using very small measurement interval values results in more problems for the original ERICA scheme than with the proposed scheme, because the proposed scheme does not measure the effective number of active connections by observing if cells are received from that connection during the measurement interval. Hence, even if the measurement interval is so short such that no cells are seen from many low-rate sources, the proposed method can compute the $FairShare$ of the bandwidth correctly.

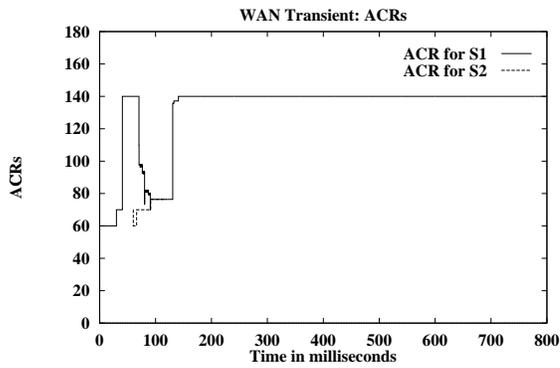 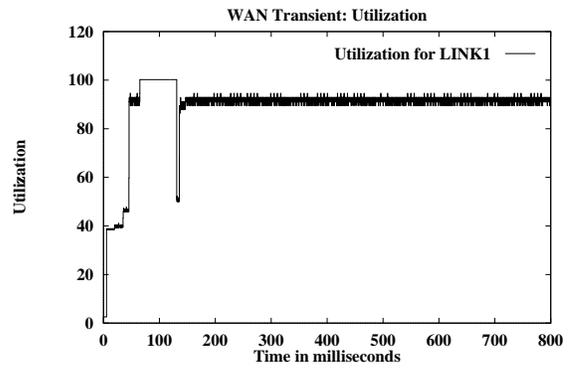

(a) Allowed Cell Rate
(b) Link Utilization

Figure 9: Results for a WAN transient configuration with ERICA

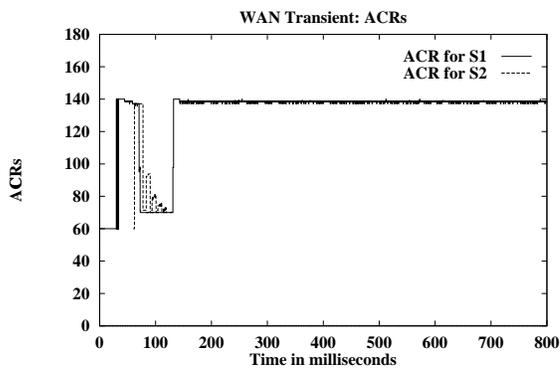 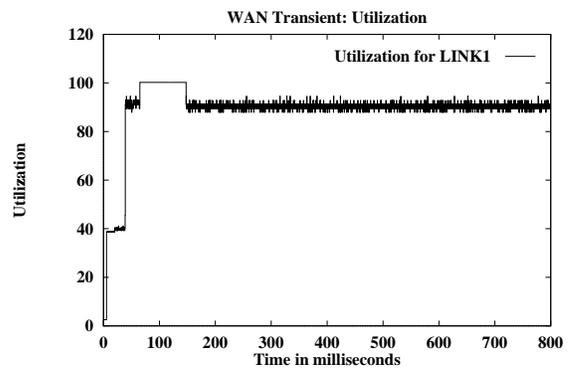

(a) Allowed Cell Rate
(b) Link Utilization

Figure 10: Results for a WAN transient configuration with the proposed ERICA and source rate measurement at the switch

- Without source rate measurement at the switch for each VC, the value of $N_{eff}$ depends on the source ACR, which is not the same as the source rate for source bottleneck cases. Thus, $N_{eff}$ is too large in those cases, and the $FairShare$ term is less than the CCR by Overload term, leading to unfairness. With per-VC source rate measurement, the value of $N_{eff}$ is correct.

## 8  Summary

This paper has proposed and demonstrated a new method to compute the fair bandwidth share for ABR connections in ATM networks. The method relies on distinguishing between underloading connections and overloading connections, and computing the value of the "effective" number of active connections, based on their activity levels. The available bandwidth is divided by this effective number of active connections to obtain the fair bandwidth share of each connection.

The method is provably max-min fair, and can be used to ensure the efficiency and fairness of bandwidth allocations. Integrating this method into ERICA tackles the fairness and measurement interval problems of ERICA, while maintaining the fast transient response, queuing delay control, and simplicity of the ERICA scheme. Analysis and simulation results were used to investigate the performance of the method. From the results, it is clear the method overcomes the fairness problem with the original ERICA, as well as its excessive sensitivity to the length of the measurement interval.

We have extended this method to include minimum rate bounds in [18]. We have also used it for point-to-multipoint connections, and noted extensions required for multipoint-to-point connections [6]. We are currently investigating using a similar load-based technique for Random Early Detection (RED) and Explicit Congestion Notification (ECN) marking in the Internet.

## Author Biographies

Sonia Fahmy is an assistant professor at the Computer Science department at Purdue University. She completed her PhD degree at the Ohio State University in August 1999. She is currently investigating multipoint communication, transport of voice and video over the Internet, and wireless network transport and applications. She has been very active in the Traffic Management working group of the ATM Forum, and has participated in several IETF working groups. Her work is published in over 40 ATM Forum contributions, and over 30 journal and conference papers. Sonia is a member of the ACM, IEEE, Phi Kappa Phi, Sigma Xi, and Upsilon Pi Epsilon, and is listed in International Who's Who in Information Technology. She received the Schlumberger foundation technical merit award in 2000 and 2001. She has served on the program committees of INFOCOM, ICNP and ICC, and co-chaired the SPIE conference on scalability and traffic control in IP networks in 2001.

Raj Jain is a Co-founder and Chief Technology Officer of Nayna Networks, Inc– an optical systems company in Milpitas, CA. He is currently on a leave of absence from Ohio State University in Columbus, Ohio, where he is a Professor of Computer and Information Sciences. He is a Fellow of IEEE, a Fellow of ACM and is on the Editorial Boards of Computer Networks: The International Journal of Computer and Telecommunications Networking, Computer Communications (UK), Journal of High Speed Networks (USA), and Mobile Networks and Applications. In the past, he was also on the Editorial Board of IEEE/ACM Transactions on Networks, was an ACM Lecturer, and an IEEE Distinguished Visitor. He is currently a Distinguished Lecturer for the IEEE Communications Society. He received a Ph.D. degree in Computer Science from Harvard in 1978 and is author of "Art of Computer Systems Performance Analysis," published by Wiley and "FDDI Handbook: High-Speed Networking with Fiber and Other Media" published by Addison Wesley. He has 14 patents and over 90 publications. For his publications, talks, and other information, please see http://www.cis.ohio-state.edu/~jain/ Raj Jain is on the Board of Technical Advisors to EdgeNet Communications Corporation, Burlingame, CA, Corona Networks, Inc., Milpitas, CA, Chip Engines, Inc., Sunnyvale, CA, SiOptic Networks, San Jose, CA, Tivre Networks, San Jose, CA, Irvine Networks, Irvine, CA, Beacon Telco, Boston, MA, and on the Board of Research Advisors to iBEAM Broadcasting Corporation, Sunnyvale, CA. Previously, he was also on the Board of Advisors to Nexabit Networks, Westboro, MA, which was acquired by Lucent Corporation.

Shivkumar Kalyanaraman is an Assistant Professor at the Department of Electrical, Computer and Systems Engineering at Rensselaer Polytechnic Institute in Troy, NY. He received a B.Tech degree from the Indian Institute of Technology, Madras, India in July 1993, followed by M.S. and Ph.D. degrees in Computer and Information Sciences at the Ohio State University in 1994 and 1997 respectively. His research interests are in the areas of traffic management, automated network management, multicast and multimedia networking. His special interest lies in the interdisciplinary areas between traffic and network management, control theory, economics, scalable simulation technologies and video compression. He is an associate member of the ACM and IEEE.

Rohit Goyal received the B.S. degree in computer science from Denison University, Granville, Ohio, and the M.S. and PhD. degrees in computer and information science from the Ohio State University, Columbus, Ohio. He is currently with Axiowave Networks, Marlborough, MA. Prior to joining Axiowave, he was a senior software engineer in the core routing division of Lucent Technologies InterNetworking Systems, formerly Nexabit Networks. His main areas of interest are MPLS, traffic engineering, QoS and traffic management. He has several journal, conference and standards publications. He is an active member of the ATM Forum and the Internet Engineering Task Force.

Bobby Vandalore received his B.Tech degree in 1993 from Indian Institute of Technology, Madras, in Computer Science. He received his MS and PhD in Computer Information Science from The Ohio State University in 1995 and 2000 respectively. He is currently a software engineer at Nokia, Mountain View, CA. His main research interests are in the areas of multimedia communications, traffic management, and perfor-mance analysis. He is the author of several papers and ATM Forum contributions. He is a member of the ACM and IEEE-CS.